\documentclass[10pt, conference]{IEEEtran}
\IEEEoverridecommandlockouts
\usepackage{cite}
\usepackage{amsmath,amssymb,amsfonts}
\usepackage{algorithmic}
\usepackage{graphicx}
\usepackage{textcomp}
\usepackage{xcolor}
\usepackage{tikz}
\usepackage{todonotes}
\usepackage{multicol}
\usepackage{hyperref}
\usepackage{booktabs}
\usepackage{multirow}
\usepackage{xcolor}
\usepackage{listings}
\usepackage{xcolor}
\usepackage[hashEnumerators,smartEllipses]{markdown}
\usepackage{etoolbox}
\makeatletter
\patchcmd{\@makecaption}
  {\scshape}
  {}
  {}
  {}
\makeatother

\definecolor{codegreen}{rgb}{0,0.6,0}
\definecolor{codegray}{rgb}{0.5,0.5,0.5}
\definecolor{codepurple}{rgb}{0.58,0,0.82}
\definecolor{backcolour}{rgb}{0.95,0.95,0.92}

\lstdefinestyle{mystyle}{
  backgroundcolor=\color{backcolour}, commentstyle=\color{codegreen},
  keywordstyle=\color{magenta},
  numberstyle=\tiny\color{codegray},
  stringstyle=\color{codepurple},
  basicstyle=\ttfamily\footnotesize,
  breakatwhitespace=false,         
  breaklines=true,                 
  captionpos=b,                    
  keepspaces=true,                 
  numbers=left,                    
  numbersep=5pt,                  
  showspaces=false,                
  showstringspaces=false,
  showtabs=false,                  
  tabsize=2
}

\lstdefinestyle{Bash}{
  language=bash,
  basicstyle=\small\sffamily,
  numbers=left,
  numberstyle=\tiny,
  numbersep=3pt,
  frame=tb,
  columns=fullflexible,
  backgroundcolor=\color{yellow!20},
  linewidth=0.9\linewidth,
  xleftmargin=0.1\linewidth
}
\def\BibTeX{{\rm B\kern-.05em{\sc i\kern-.025em b}\kern-.08em
    T\kern-.1667em\lower.7ex\hbox{E}\kern-.125emX}}
\begin{document}

\title{ParticleGrid: Enabling Deep Learning using 3D Representation of Materials
{\footnotesize \textsuperscript{}}
\thanks{\textsuperscript{*}Equal contribution}
}

\author{\IEEEauthorblockN{Shehtab Zaman*} 
\IEEEauthorblockA{Binghamton University\\
  Binghamton, New York, USA\\
  szaman5@binghamton.edu}
  \and
\IEEEauthorblockN{Ethan Ferguson*}
\IEEEauthorblockA{Binghamton University\\
  Binghamton, New York, USA\\
  efergus3@binghamton.edu}
 \and
\IEEEauthorblockN{Cécile Pereira}
\IEEEauthorblockA{TotalEnergies Marketing and Services\\
  Paris, France\\
  cecile.pereira@totalenergies.com}
\and
\IEEEauthorblockN{Denis Akhiyarov}
\IEEEauthorblockA{TotalEnergies EP Research \& Technologies US\\
 Houston, Texas, USA\\
 denis.akhiyarov@totalenergies.com}
\and
\IEEEauthorblockN{Mauricio Araya-Polo}
\IEEEauthorblockA{TotalEnergies EP Research \& Technologies US\\
 Houston, Texas, USA\\
 mauricio.araya@totalenergies.com}
\and
\IEEEauthorblockN{Kenneth Chiu}
\IEEEauthorblockA{Binghamton University\\
 Binghamton, New York, USA\\
 kchiu@binghamton.edu
}
}


\maketitle

\begin{abstract}
From AlexNet to Inception, autoencoders to diffusion models, the development of novel and powerful deep learning models and learning algorithms has proceeded at breakneck speeds.
In part, we believe that rapid iteration of model architecture and learning techniques by a large community of researchers over a 
common representation of the underlying entities has resulted in transferable deep learning knowledge. As a result, model scale, accuracy, fidelity, and compute performance have dramatically increased in computer vision and natural language processing. On the other hand, the lack of a common representation for chemical structure has hampered similar progress.
To enable transferable deep learning, we identify the need for a robust 3-dimensional representation of materials such as molecules and crystals. The goal is to enable both materials property prediction and materials generation with 3D structures.
While computationally costly, such representations can model a large set of chemical structures. We propose \textit{ParticleGrid}, a SIMD-optimized library for 3D structures, that is designed for deep learning applications and to seamlessly integrate with deep learning frameworks. Our highly optimized grid generation allows for generating grids on the fly on the CPU, reducing storage and GPU compute and memory requirements.
We show the efficacy of 3D grids generated via \textit{ParticleGrid} and accurately predict molecular energy properties using a 3D convolutional neural network. Our model is able to get 0.006 mean square error and nearly match the values calculated using computationally costly density functional theory at a fraction of the time. 


\end{abstract}

\begin{IEEEkeywords}
Deep Learning, Scientific Machine Learning, High-Performance Computing, 3D Structures
\end{IEEEkeywords}
\IEEEpeerreviewmaketitle

\section{Introduction [1 pg]}

Deep learning (DL) has had a transformative effect on disciplines from natural language processing, computer vision to reinforcement learning. It is expanding the way we perform science. Whether it is news articles generated by GPT-3\cite{gpt3}, paintings in the style of Van Gogh by StyleGAN~\cite{stylegan, stylegan2}, or conditional images generated by Dall-E \cite{dalle}, generative modeling is one of the most striking uses of DL. 

The aforementioned models achieve such amazing results by building on the work of a wide community of researchers consistently innovating on DL algorithms. For image classification, iteration over convolutional neural networks (CNNs) resulted in rapid improvements in both predictive and computing performance from AlexNet to ResNet and InceptionNet to EfficientNets. Similarly, for image segmentation we've seen improvements from R-CNN, Faster R-CNN, to YOLO, allowing for real-time image segmentation. We have seen similar trends in image generation with Generative Adversarial Networks (GANs), Variational Autoencoders (VAEs), Normalizing Flows, and Diffusion models. Due to the standardization of input, we see a trend of rapid discovery of models, layers, and learning techniques to facilitate performance improvements. Images have a natural and well-established representation. As a result, we've seen significant improvements in DL-based tasks in a relatively short amount of time.

\begin{figure*}
    \centering
    \includegraphics[width=\linewidth]{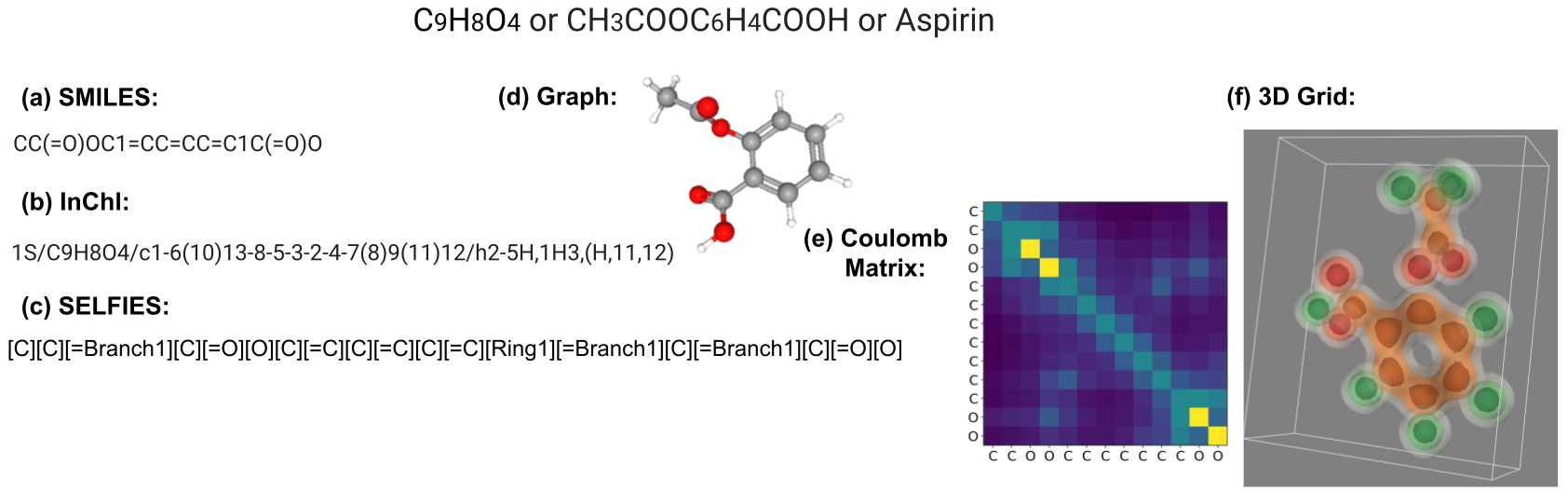}
     \caption{The many machine-readable views of the  molecule $C_9H_8O_4$, commonly known as Aspirin. \textbf{(a)} is the canonical SMILES retrieved from the PubChem database. \textbf{(b)} is the INchI (International Chemical Identifier) representation from the same  database. \textbf{(c)} is the SELFIES  string obtained from Krenn et al. \cite{selfies}. The graph-like representation in \textbf{(d)} is visualized using the ball-and-stick model. Each atom corresponds to a vertex, and each bond corresponds to an edge in the graph.  
     \textbf{(e)} \textbf{(f)} is the visualization of a \textit{ParticleGrid} generated grid using Mayavi \cite{mayavi}.}
    \label{fig:representations}
\end{figure*}

Yet, DL-based generative models are still at a nascent stage for scientific principles. Applications in molecular and materials discovery have yet to be utilized in wide applications. Generative models of chemical structures especially have yet to experience similar growth as in image and sequence generation. This can be largely traced to the difficulty in identifying a common representation for chemical structures. Molecules and materials cover a large set of possible combinations of atomic elements with varying structures, and varying representations, each with its own strengths and drawbacks. Some classes of materials such as molecules have textual representations such as SMILES \cite{1988smiles} and SELFIES \cite{selfies}. The same molecules are also succinctly represented using graphs and graph-based structures such as coulomb matrices ~\cite{coulomb, coulomb-eig}. Materials such as crystals on the other hand do not share such representations. The difficulty in the representation of these materials is further compounded due to the spatial symmetry requirements and periodicity. We can see the various representations of molecular structures in Fig. \ref{fig:representations}.

In recent years, due to larger GPU memory and faster compute capabilities, three-dimensional grid-like representations have been explored. The importance of three-dimensional structural information can be highlighted as large-scale datasets such as the
open graph benchmark - large-scale challenge (OGB-LSC)  \cite{hu2021ogb} updated their datasets to include 3D coordinate information. This follows the trend as \cite{bolton2011pubchem3d} employed 3D conformer models to extend PubChem \cite{pubchem}, an open repository for small molecules designed to study biological activity. 3D grids can encode both geometric and chemical information, while also being generalizable to  represent many different classes of materials. Including geometric and chemical information to the input of deep learning models provides inductive bias to improve learning results ~\cite{griffiths2010probabilistic, battaglia2018relational}. Representations that can be shared across many different classes enable the use of transfer learning \cite{zhuang2020comprehensive}. Significantly, this enables the use of deep learning, traditionally requiring large datasets, to be used on classes of materials with few samples by pre-training on larger datasets. 

3D generative models as in \cite{zeogan, ragoza2020learning} show that there are promising DL models that can be utilized to augment molecule and crystal generation workflows. The ability to represent different classes of materials such as organic molecules and crystals shows that grid-like representations may be promising as a representation for DL tasks. As grids are significantly larger than more concise objects such as SMILES, and graphs, the computation and storage costs can be prohibitive. Therefore we propose, \textit{ParticleGrid}, a SIMD-optimized grid generation for molecules and crystals. Our goal is to provide on the fly grid generation abilities from 3D coordinates and seamlessly integrate with data ingestion pipelines of deep learning frameworks such as PyTorch and Tensorflow. We also propose a general-purpose transformation from grids to 3D coordinates. 

We discuss the various strengths and shortcomings of text-based, graph-based, and grid-based representations in \textbf{Section II}. In \textbf{Section III}, we present the mathematical foundation of our grid representations. In \textbf{Section IV}, we show the reversibility of the grid to discrete coordinates. \textbf{Section V} and \textbf{Section VI} show the programming interface and detail the optimizations in \textit{ParticleGrid}, respectively. \textbf{Section VII}, we show an application of 3D structures for molecular property prediction, and finally, \textbf{Section VIII}, discusses the results of our findings. 

We publicly release our implementation as an open-source library, and the performance and application codes can be found for reproducibility at \url{www.github.com/ParticleGrid/ParticleGrid}

\section{Related Work}

Deep learning algorithms have been used in materials science applications to both perform supervised prediction of material properties such as energy, bond forces, and others, and also the unsupervised generation and reinforcement learning of new molecules and crystals for drug discovery and delivery, carbon capture and storage, and batteries among others.
This work has used a variety of different representations for materials.
Broadly, we categorize them as string-based, graph-based, and grid-based.

Gomez-Bombarelli et al.~\cite{gomez2018automatic} use deep generative models to generate new SMILES (Simplified Molecular Input Line Entry Specification) strings.
Kusner et al.~\cite{grammarsmiles} extend such generative algorithms by imposing syntactic and semantic constraints on the decoder.
While generating valid structures,
the SMILES representation is inherently unable to capture many of the physical invariances present in molecules 
\cite{guimaraes2017objective}.
Furthermore, similar structures may have markedly different SMILES representations.
In order to mitigate the structural representation limitations of SMILES, recently the SELFIES (SELF-referencIng Embedded Strings) representation has been proposed \cite{selfies}. Using formal grammar, Krenn et al. derive a representation that is highly robust. Even a random arrangement of the symbols defined in the grammar result in a valid molecule. This is a significant characteristic, as generative models benefit greatly from such a robust representation. 
While using string representations is very attractive for generative modeling, they suffer from some drawbacks. 
Other classes of materials such as periodic crystals also pose difficulties for SMILES and SELFIES.
Whilst future work can enable these character-based representations for a wider variety of materials, currently, these representations are limited to molecules \cite{krenn2022selfies}.
%
Yao et al. developed RFCode, a string representation of crystals based on SMILES and InChI for reticular frameworks such as metal-organic frameworks (MOFs). Reticular frameworks are nanoporous crystals with many constituent atoms and use in diverse fields such as gas storage, separation, and catalysis \cite{yao2021inverse}. The authors designed a data-driven inverse design framework to propose new MOFs for optimized methane storage using a VAE. 

Graph-based representations of molecules have also been used for molecular generation tasks. Jin et al.~\cite{junctiontree} use graph-structured encoder and junction tree-based decoder to generate molecular structures. While graph-structured generative models allow for greater representation of long-range atomic interactions and provide geometric information of the molecule, graph-structured generative models have been restricted to a small and limited set of molecules \cite{de2018molgan}. On the other hand. MPNN, CGCNN, SchNet, and GemNet, achieve state-of-the-art accuracy in molecular property prediction, and molecular dynamics are achieved by graph-based neural networks \cite{mpnn, cgcnn, schnet, gemnet}. 
Yet, these models are not suitable for use in unsupervised training and generative modeling. Graph-based generators are required to sample discrete structures, i.e graphs, from a continuous latent space. This results in poor performance when the discrete structure is too large \cite{de2018molgan}. 

Grid-based modeling of materials can provide both geometric information and generalizability. Hoffman et al. explore the possibility of performing dimensionality reduction using deep learning models to encode and decode 3D crystal structures \cite{hoffmann2019data}. Ragoza et al.~\cite{ragoza2020learning} use grids with Gaussian-like density functions for voxels to learn continuous representations of 3D molecular conformations. Our formalism builds on this by allowing for grids with variable spatial resolutions. Each grid can represent a different volume of space depending on the size of the material. This enables the model to learn multi-resolution representations of molecules. This allows us to represent different classes of materials that may have different numbers of atoms required  to scale. As we integrate the atomic probability densities over voxel regions, we are also able to encode global molecule level features such as the total number of atoms, and the number of atoms per element. Furthermore, the discretization step described by Ragoza et al. utilizes a solver that uses gradient descent for atomic structure detection. They generate multiple possible structures per grid. They subsequently use beam search to find the best possible 3D structure. Similarly, Kim et al. use a 3D-GAN  to generate crystal structures called, Zeolites \cite{zeogan}.


A significant issue with using molecular grids is the compute and memory requirements. Deep learning models require high data throughput, especially in high-performance setting. Generating grids and storing them on disks is infeasible, as datasets can be scaled up to millions and billions of samples. As a result, the grids must be generated on the fly. Yet, high-performance vectorized libraries such as Numpy \cite{harris2020array} cannot provide sufficient throughput. Therefore a more low-level implementation is required, such as \textit{ParticleGrid}. We provide direct comparisons with the grid generation times in Section VI.


\section{Grid Representation of Molecules}

We define three objectives for a suitable grid representation for deep learning (DL) for material science: 

\begin{enumerate}
    \item Physically inspired
    \item Generalizable
    \item Invertible
\end{enumerate}

A suitable grid needs to provide physically relevant information not only to provide suitable inductive bias ~\cite{battaglia2018relational, griffiths2010probabilistic}, but also to provide important information to practitioners. Inductive biases can be physical laws, assumptions, or human knowledge that can be inscribed into the data, model, and loss function to prioritize a specific solution and improve learning performance. Visually inspecting the grid should provide significant information to a scientist. The generalizability criterion is significant for using the grid for DL. As we have seen before, DL models using the same representation allow multiple different types of models to be used in conjunction. Also significantly, re-using learned models in transfer learning is imperative in low-data settings. Data availability in scientific research can be bimodal. There exist large datasets with millions and billions of samples as is the case for small organic and drug-like molecules. On the other hand novel, exotic materials may only have a hundred or thousand known samples. Therefore, a general representation that can cover many subdomains of material science is important. Finally, the invertibility of the representation is the ability to transform the structure to other canonical representations. First principle methods have long-established algorithms to quantify and classify materials. DL methods, especially generative models, are designed to augment the established scientific discovery pipeline. Therefore the data representation requires interoperability between representations suited for DL and representations required by other computational and experimental methods.  

\subsection{Voxel-filling Function}
With the aforementioned criterion in mind, we propose a grid representation discretizing the real space realization of an atom in to grids. 

A simple realization of our approach is a grid representing a single molecule is a 4-dimensional tensor of the form (Channel, H, W, D). Each channel of the data cube represents the density of the element at the point. Generally, we use a cube grid, there for H=W=D, but not required. Each voxel represents   $\left(\frac{1}{N}\right)^3 (\text{ABC})$ volume of 3D space. $N$ is the number of voxels, usually $N = H\times W \times D$. Here, A, B, and C represent the height, width, and depth of a bounding box of the molecule. The user has control over choosing the orientation and size of the bounding box. It is important to note that the voxel need not be cubic, and the grids can have different resolutions between molecules.

For a molecule at point $\vec{\mu}= (\mu_x,\mu_y,\mu_z)$, the function defining the density of the atom existing at point $\vec{p} = (x, y, z)$ is given by $f_{\vec{\mu}}(\vec{p})$, the 3D Gaussian probability distribution function.

\begin{equation} \label{Eq:Multi_Gaussian}
    f_{\vec{\mu}}(\vec{p}) = \frac{1}{\sigma^3 \left(2 \pi \right)^{\frac{3}{2}}} e^{ -\frac{d(\vec(\mu),\vec{p})}{2\sigma^2}} 
\end{equation}

Where, $ d(\vec(\mu),\vec{p})$, is the element-wise distance of the functions of the inputs. For a non-periodic molecule, we use the squared euclidean distance. For the case of periodic crystals, the nearest image distance may be used. The nearest image distance of an atom on a periodic lattice is $\text{min}(d(\mu^{i}_x, x_0))$ where, $ \mu^{i}_x \in T(\mu_x)$, where $T(\mu_x)$ is the set of coordinates of the lattice vector translations. $\sigma$ is the variance or the "spread" of the Gaussian. 

Gaussians are a natural choice for a function to use for spatial representation molecules. %
Furthermore, since we used a correctly normalized 
the sum of the grid values for all the channels should equal the number of atoms, and the sum over each channel should equal the number of atoms of that type. 

For a given atom at position $(\mu_x,\mu_y,\mu_z)$, on channel q, the value of the the region $\mathcal{D}^{q}_{i,j,k} = [(x_i,y_j,z_k),(x_i+\delta_x,y_j+\delta_y,z_k+\delta_z)]$ is defined by Eq. \ref{Gridding_func}. Here, the $\mathcal{D}^{q}_{i,j,k}$ is the value of the $i,j,k$-th voxel. $x_i,y_j,$ and $z_k$ are the left endpoints of the real space locations of the $i,j,k$-th voxel. $\delta_x, \delta_y,$ and $\delta_z$ are the size of side of the voxels at their respective dimensions. 





\begin{align}\label{Gridding_func}
 F_{\vec{\mu}}(\vec{p}) &= \left(\frac{1}{2}\right)^{3} G_{\mu_x}(x_a)H_{\mu_y}(y_b)L_{\mu_z}(z_c) \\
 \label{eq:x_integral}
 G_{\mu_x}(x) &= \mathrm{erf} \left(\frac{\sqrt{2}}{2\sigma}\left(\mu_i - x \right)\right)\biggr\rvert^{x_i+\delta_x}_{x_i}\\
  \label{eq:y_integral}
 H_{\mu_y}(y) &= \mathrm{erf} \left(\frac{\sqrt{2}}{2\sigma}\left(\mu_j - y \right)\right)\biggr\rvert^{y_j+\delta_y}_{y_j}\\ 
  \label{eq:z_integral}
 L_{\mu_z}(z) &= \mathrm{erf} \left(\frac{\sqrt{2}}{2\sigma}\left(\mu_k - z \right)\right)\biggr\rvert^{z_k+\delta_z}_{z_k}
\end{align}

Since the Gaussian in Eq. \ref{Eq:Multi_Gaussian} is separable, we can separate the integral evaluated in Eq. \ref{Gridding_func} into three functions Eq \ref{eq:x_integral}, \ref{eq:y_integral}, and \ref{eq:z_integral}.

Given a molecule as a coordinate set, $\mathcal{M}$, such that each molecule at coordinate $\vec{\mu} = (\mu_x,\mu_j,\mu_k)$ is in set $\mathcal{M}$.
The grid is generated by evaluating $F_{\vec{\mu}}(\vec{p})$ over all $(H \times W \times D )$ regions for all molecules $(\mu_i, \mu_j, \mu_k) \in \mathcal{M}$. 


We can also see that due to this formulation, we have the condition, 

\begin{equation} 
\sum^{\text{channel}}_{q}\sum^N_{i,j,k} \mathcal{D}^{q}_{i,j,k} = \text{Number of Atoms} 
\end{equation}

It is important to note the voxel-filling process does not require a fixed bounding and therefore resolution. Similar to images, the same molecule can have multiple grid representations with differing resolutions. This allows us to use the same sized grid for molecules with varying volumes and numbers of elements.

Equation \ref{Gridding_func}. can be extended to augmented to produce weighted grids. Also, other grid-like structures may be used in conjunction with the probability grids such as the potential energy grids derived from the Lennard-Jones potentials used in \cite{zeogan}.  

\section{Maximum Likelihood Estimation of Atomic Coordinates}

\begin{figure*}[!ht]
    \centering
    \includegraphics[width=\linewidth]{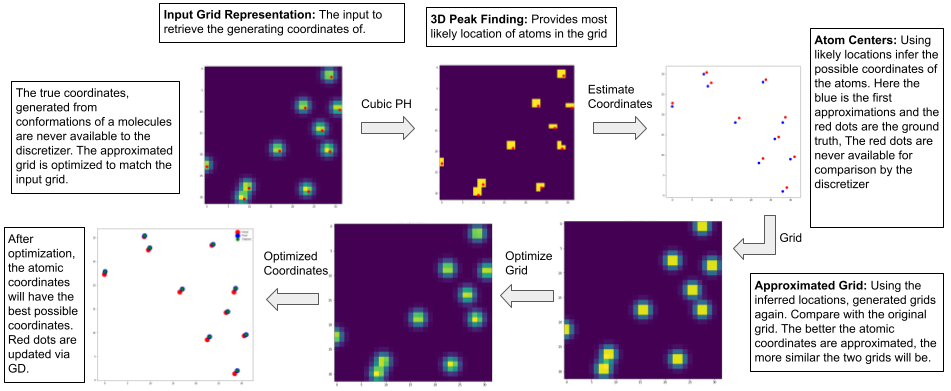}
    \caption{Reversing the gridding process to obtain molecular coordinates. We use a 2D example for ease of visualization.}
    \label{fig:discretizaion}
\end{figure*}

In order for the grids to be useful for generative modeling, the grids must be invertible to retrieve atomic coordinates. 
Usually, generative models work on grid space and have grids as inputs and outputs. It is desirable to obtain the coordinates from the grids for further processing with traditional cheminformatics and other ML methods. This allows \textit{ParticleGrid} to be a middle layer that is interoperable with string and graph-based representations and can work in conjunction with the current state of the practice.

\subsection{Algorithm for grid reversal}

The deduction of coordinates follow a two- stage process, as seen in Fig. \ref{fig:discretizaion}. 

For a given point $\mathbf{P}_0$ with coordinates  $(x_0,y_0,z_0)$ generating grid $\mathbf{G}$, we wish find $\Theta$, with coordinates $(x^\prime_0,y^\prime_0,z^\prime_0)$,  such that  $\Theta \approx \mathbf{P}_0$.
At any given time, only the original grid $\mathbf{G}$ is available.

\begin{enumerate}
    \item Approximate $\Theta$ from $\mathbf{G}$ 
    \item Generated approximated grid $\mathbf{G}^\prime$ using Eq \ref{Gridding_func} with $\Theta$ as the input
    \item Calculate the difference, $\mathcal{L}$, using mean squared error between $\mathbf{G}$ and $\mathbf{G}^\prime$
   
    \item Update $\Theta \leftarrow \Theta - \lambda\frac{\partial \mathcal{L}}{\partial \Theta}$, where $\lambda$ is the stochastic gradient descent step size (learning rate)
    
    \item Repeat steps 2-4 until $\mathcal{L} < \epsilon$, for some small value of $\epsilon \in \mathbf{R}$
\end{enumerate}

\subsection{Optimal initialization of approximate coordinates}

As a result of our grid structure, atomic centers display a clearly defined local maxima nearest to the atomic centers. Reversing the grid to obtain possible coordinates of generated grids constitutes a maximum likelihood estimation of the local maxima across the grid. This requires mode detection on the 3D grids. While there are many possible ways to estimate peaks on a 3D volume, we explore the use of persistent homology \cite{ph-review}. Persistent homology is a topological data analysis (TDA) tool to extract information from simplicial complexes. Defining the 3D grids as 3D cubical complex, we know the topologically significant structures will be atomic centers. We use the efficient CubicalRipser library \cite{kaji2020cubical} to compute 2-cycles on our 3D grids by computing the cohomology by matrix reduction. For further details on the algorithm refer to \cite{de2011dualities}. 

\subsection{Optimizing the coordinates}

After obtaining the approximate coordinate, $\Theta$, we can use stochastic gradient descent to optimize the approximated coordinates to match the original grid. 

Therefore, we need to be able to calculate the gradient update value $\frac{\partial \mathcal{L}}{\partial \Theta}$ efficiently to update the our initial set of coordinates $\Theta$. 

We know the loss, given positions $\Theta $ or $(x,y,z)$ is

\begin{align}
    \mathcal{L} &= ||\mathbf{G} -\mathbf{G}^\prime ||^2_2 \\
    \mathcal{L} &= \sum_{i,j,k}|| \mathbf{G} - F_{\Theta}(x_i,y_j,z_k )||^2_2
\end{align}

Here, $\mathbf{G}^\prime$ is the grid generated using the approximated  
For simplicity of notation, we use $F_{ijk}$ to refer to the function $F_{\Theta}(x_i,y_j,z_k )$ from here on. 

Updates to $\Theta$ can be separately achieved for each coordinate so we have the set of equations, 

\begin{align}
    x^{t+1}_0 &\leftarrow x^t_0 - \lambda \frac{\partial \mathcal{L}}{\partial x^{t}_0}
\end{align}
So the partial derivative with respect to the coordinate $x_0$ is given by 
\begin{align}
    \frac{\partial \mathcal{L}}{\partial x_0} &= \sum_{i,j,k}
        \frac{\partial \mathcal{L}}{\partial F_{ijk}}
        \frac{\partial F_{ijk}}{\partial x_0} \\
    &= \sum_{i,j,k}2\left( F_{ijk} - G_{ijk}\right) \frac{\partial F_{ijk}}{\partial x_0}
\end{align}

At this point, we can use the Fundamental theorem of calculus to evaluate $\frac{ \partial F_{ijk}}{\partial x}$. 

It is important to note that Eq. \ref{Eq:Multi_Gaussian} and Eq. \ref{Gridding_func} are separable in x, y, and z dimensions. 
We know that for some separable multivariate function 

\begin{align}
    F &= G(x)H(y)P(z) \\
    \frac{\partial F}{\partial x} &=H(y)P(z)\frac{\partial G}{\partial x}
\end{align}

and $F = \int f dx$ then $\frac{dF}{dx} = f$. So, the partial derivative term for the x coordinate is, 

\begin{equation}
    \frac{\partial F_{ijk}}{\partial x_0} = \left(\frac{1}{8} H_j(y_0)L_k(z_0)\right) \frac{\partial G_i(x_0)}{\partial x_0}
\end{equation}

Since we know that, 

\begin{equation}
    \frac{d}{dt}\mathrm{erf}(t) = \frac{2}{\sqrt{\pi}}e^{t^2} 
\end{equation}

We have, 

\begin{equation}
    \frac{\partial G_i(x_0)}{\partial x_0} = \frac{\sqrt{2}}{\sigma\sqrt{\pi}} \left(
    e^{-\frac{(i-x_0)^2}{2\sigma^2}}- e^{-\frac{(i+\delta_x-x_0)^2}{2\sigma^2}}
    \right)
\end{equation}
The other derivatives follow the same pattern. 

\section{ParticleGrid's Programming Interface}

\textit{ParticleGrid}, is designed to enable SciML\footnote{SciML: Scientific Machine Learning} practitioners to quickly and easily generate grids for molecular structures. In order to minimize the friction of usage, a high-level interface developed in Python binds to a C++ backend. The entire library can be installed using the Python package manager Pip, with command: 

\begin{lstlisting}[caption=Installation instructions]
pip install git+https://github.com/ParticleGrid/ParticleGrid.git
\end{lstlisting}

We ensure low installation overhead by only requiring a C++17 enabled compiler and the Numpy \cite{harris2020array}package.

One of the major considerations for \textit{ParticleGrid} is to smoothly integrate with deep learning frameworks such as PyTorch \cite{NEURIPS2019_9015} and TensorFlow \cite{tensorflow2015-whitepaper}. We designed our implementation to work seamlessly with the data loading mechanism for these frameworks, and therefore optimized for CPU performance, in order to generate grids on background processes while training on accelerators. Also, Numpy arrays holding grids are produced, in order to facilitate zero-copy transfers to PyTorch and TensorFlow data containers. A simple grid generation sample can be found in Listing \ref{code:grid-gen}.

\begin{lstlisting}[language=Python, caption=Grid generation example, label=code:grid-gen]
import numpy as np
import torch
import tensorflow as tf
from ParticleGrid import coord_to_grid

# Points are in the format (channel, x, y, z)
test_points = np.array([0, 0.5, 0.5, 0.5],
                       [1, 0.0, 0.1, 0.2])
# Generates a (2,32,32,32) grid
grid = coord_to_grid(test_points,
                     width=1,
                     height=1,
                     depth=1,
                     num_channels=2,
                     grid_size=32,
                     variance=0.05)

# Convert to PyTorch tensor
grid_torch_tensor = torch.from_numpy(grid)
# Convert to TensorFlow tensor
grid_tf_tensor = tf.convert_to_tensor(grid)
\end{lstlisting}

Similarly, we also provide an easy and extendable workflow to reverse the grid. In \textbf{Section V.B} and in Fig. \ref{fig:discretizaion} we use the Cubical Ripser library in order to use persistence homology to determine voxels with atoms. 

\section{Grid Generation Implementation}
The following development and testing are performed on an Intel Xeon Silver 4114 CPU @ 2.20 Ghz, with 32k L1i and L1d cache, 1024K L2 cache, and 14080k L3 cache, 128 GB of system RAM. We used the GCC 7.5.0 compiler. 

\subsection{Baseline}
Because the value at each grid point determined by Eq. \ref{Gridding_func}.
is separable into $G$, $H$, and $L$,  
these values can be computed and stored for their respective axes ($x$, $y$, and $z$) to later compute the value
at a given grid point.
This already is a significant speedup over the naive approach of calculating $F_{i,j,k}(x,y,z)$ at each grid point.
Furthermore, because each part of the separable function is the difference of a function calculated at two points, 
and the minuend of one calculation is the subtrahend of the next, the value of the \textbf{erf} can be calculated half as many times.
These optimizations provide much faster algorithmic performance over a naive baseline implementation, 
but the performance still leaves much to be desired, 
and exploiting the fact that much of the grid can be effectively ignored 
provides fewer memory accesses and computations without degrading accuracy.

\subsection{Exploiting Truncation}

Since each of Eq. \ref{eq:x_integral}, \ref{eq:y_integral}, and \ref{eq:z_integral} rapidly approach zero when calculated at increasing distance from an atom.
Because $F = G(x)H(y)L(z)$, if the values of $G$, $H$, or $L$ are sufficiently small, they can be ignored.
And, where these functions are insignificant can be known before calculating $F(x, y, z)$ for each point by simply searching through all $x$ for $G(x)$, $y$ for $H(y)$, and $z$ for $L(z)$.
The search space for these functions grows with $N$, whereas the number of points to be calculated grows with $N^3$, where $N$ is the size of the grid. 
So, it is very economical to add a limited computation ahead of time to be able to skip many calculations of $F$.
In fact, we found through performance profiling that over 99\% of the time taken to compute a grid was in the multiplication of the three functions, 
so any optimizations there will be greatly beneficial.
Not only can many calculations of $F$ be skipped, but so can writing all of the zero-valued points of $F$,
because the tensor representing the grid is initialized to zero.

How quickly the functions shrink depends on the variance, but the variance can be tuned so that the falloff is over any desired distance. In our testing (See \ref{sec:evaluation}), we found that the specific variance chosen is relatively inconsequential so long as it is ``reasonable'', as can be seen in \ref{fig:sparse}. That is, as long as individual atoms are readily distinguished, the speedup while exploiting sparsity stays approximately the same.
This is particularly useful because with floating point numbers, the value of these functions rapidly saturates to zero in practice.
For example, using 32-bit floating point numbers, a variance of 0.5, 
and a grid size of 64 on a group of 100 molecules from the MOSES dataset \cite{MOSES}
showed that out of the total $64^3=262144$ cells in each grid, an average of only $2623$ cells were non-zero. That is $1\%$ of all cells, allowing for significant skipping of calculation and memory writes. 

As suggested before, saturation to zero is not necessary but it is convenient. A threshold could be used instead of relying on floating point imprecision. 
In this case, instead of seeing where $G$, $H$, and $P$ are zero before calculating $F$, we instead see where they are below some threshold, 
and do not bother doing the multiplication beyond that threshold.
However, we did not implement this because allowing floating point values to naturally saturate to zero is very effective, and largely maintains precision. Of the various optimization strategies used, 
this was the most significant. On average, we see about a 7x speedup due to this strategy in our testing,
using many different grid sizes and variances.
Specific speedup can be seen in Fig. \ref{fig:sparse}.
The speedup varies quite significantly, particularly due to grid size.
The greatest speedup is seen with $32^3$ grids.
This is likely because they have a good mix of being able to fit in cache
and benefit significantly from sparsity. $16^3$ grids do not benefit as much from sparsity as calculating a $16^3$ grid row requires accessing many cache lines representing a grid, a cache line of 64 bytes holding 8 floats is already half of a full grid row. And, non-sparse accesses all cache lines in order, which is easy for processors to predict. On the other hand, the larger grids require significant data movement. For example, a single $64^3$ grid is 1MB, thus performing calculations on larger grids with many channels easily saturates the cache and compute time is dominated by memory access.

However, there is still significant hardware speedup left on the table which can be used to further enhance efficiency.
\begin{figure}
    \hspace{10pt}
    \centering
    \includegraphics[width=\linewidth]{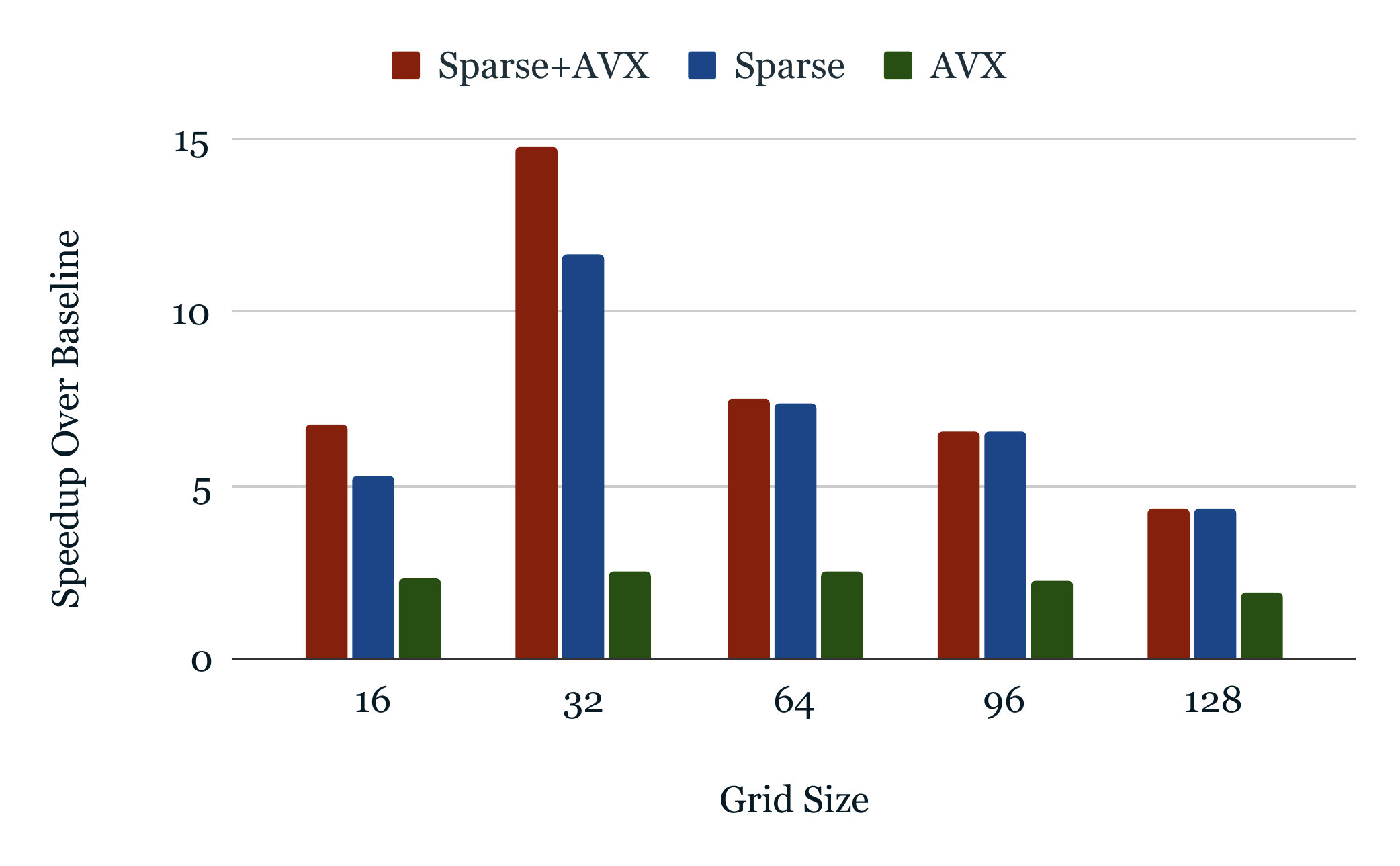}
    \caption{Speedup over baseline C++ implementation using Sparsity, AVX, and both Sparsity + AVX. }
    \label{fig:sparse}
\end{figure}

\begin{figure}
    \centering
    \includegraphics[width=\linewidth]{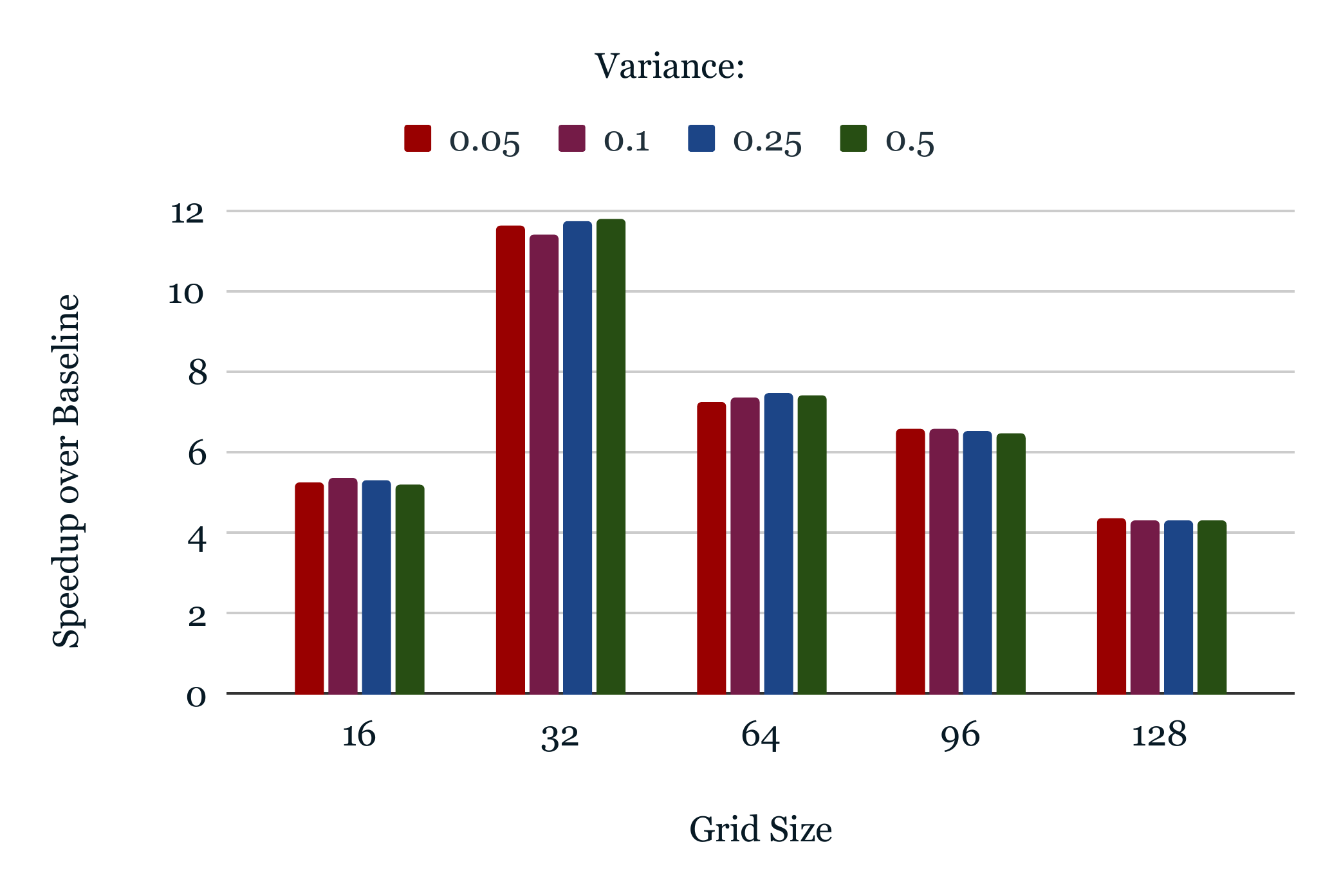}
    \caption{Speedup over baseline C++ implementation using Sparsity with different variances. The higher variance values result in increased number of non-zero voxels.}
    \label{fig:sparse}
\end{figure}

\subsection{Instructions-level parallelization of main functions}

Grid generation is a prime candidate for advanced math instructions that compute multiple values at the same time (SIMD). The functions $G$, $H$, $L$, and $F$  are readily SIMDizable. 
We added SIMD intrinsics to ParticleGrid in order to force the compiler to use AVX instructions. As can be seen when looking at the speedup of AVX over baseline, and the speedup of AVX+Sparsity over only Sparsity (See \ref{fig:sparse}), the manual addition has a significant effect beyond optimizations the compiler did automatically.

Unfortunately,
there is no Gauss error function ($\textbf{erf}$) intrinsic on an 8-float vector
in any of Intel's AVX iterations (as far as we know),
meaning there is no built-in instruction to compute $\textbf{erf}$ efficiently.
Fortunately, the Bürmann series \cite{schopf2014burmann} can be used to approximate it with little error. We use the first three terms in our approximation with $c_1 = \frac{31}{200}$ and $c_2 = \frac{341}{8000}$
This is also used in our baseline implementation.

Other than this, it is a simple case of calculating the bounds for $G$, $H$, and $L$ in a SIMD form, and doing the same for the $i$ dimension of $F$.

\subsection{Performance Evaluation}
\label{sec:evaluation}

\begin{figure}
    \centering
    \includegraphics[width=\linewidth]{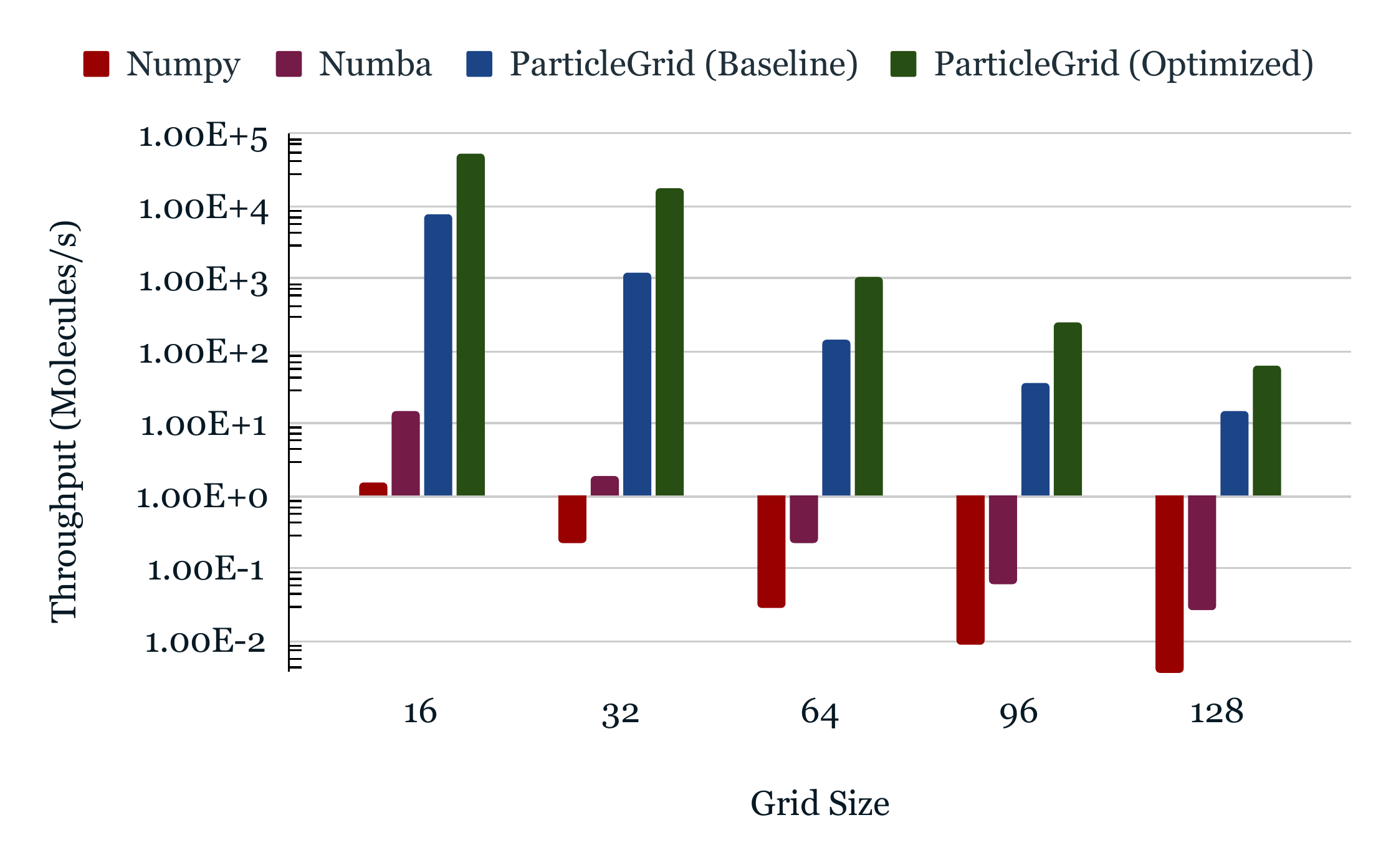}
    \caption{Throughput using NumPy, Numba, and ParticleGrid for calculating molecules, 
        in Molecules/s}
    \label{fig:throughput}
\end{figure}
In order to test performance, we use the MOSES dataset
and conformed the samples to get a set of atom positions using RDKit.
Then, we took a random sample of 100 of the resulting molecules and
conduct experiments with our NumPy, Numba \cite{lam2015numba}, Base ParticleGrid, and Optimized ParticleGrid implementations
on each molecule in the set, repeating the calculations 10 times for each variance we test, keeping track of the total time to do the calculations.
We tested the variances of 0.05, 0.1, 0.25, and 0.5, 
since they seem to represent visually acceptable results when inspecting ParticleGrid outputs.
This results in each grid generating function being run 4,000 times for each grid size.
Then we calculate the average time to compute a molecule and its inverse, the throughput.
The following is an analysis of the times we observed from these runs.

The performance impact of hardware optimizations is less than expected.
Nevertheless, the overall performance gain from using C++ and avoiding unnecessary computation is significant (See \ref{fig:throughput}).
In the most recent iteration of ParticleGrid, enabling vector instructions with sparsity already enabled only resulted in an average speedup of less than 2\% in our tests for grids with $N \geq 64$.
Smaller grids do see a more significant speedup, and enabling vector instructions without sparsity shows an over 2x speedup.
However, a speedup of 8x would be expected from a perfect vectorization implementation.
This suggests either a problem with the implementation or a hardware limitation, further analysis is underway.
The implementation is both simple and works correctly, 
so it would seem the implementation runs into a hardware bottleneck of some kind.
We believe \textit{ParticleGrid} is memory bound on our system, 
and so will minimally benefit from further hardware optimization.
However, more research and profiling is needed to pinpoint the cause.
The overall speedup from a baseline implementation can still be over 15x despite these limitations,
thanks mostly to the benefit of sparsity.
The exact testing parameters change this significantly, 
and it appears cache pollution causes very large fluctuations in speedup, 
as much as 10x in our testing when repeatedly calculating a grid for only one molecule,
instead of calculating a grid for many. 
This would also point to a memory bottleneck,
since a compute-bound process would be unlikely to see such a significant impact from memory bandwidth.
There is certainly more research to be done in this area to further identify the exact causes of various performance changes, but even without this ParticleGrid outperforms other implementations.

We originally implemented the grid generation in Python using NumPy, and also using Numba.
Compared to our optimized C++ implementations, these implementations are as much as 4 orders of magnitude slower (See \ref{fig:throughput}).
Numba gives a faster result than NumPy, but is still routinely 3 orders of magnitude slower than the C++ ParticleGrid implementation.
All of these implementations are available in the ParticleGrid repository.

One final consideration is in the context of machine learning, many molecules will be needed in quick succession.
Because separate molecules in the same batch can all be learned on in parallel,
an ML framework can do grid computation for all of them in parallel.
Commonly used frameworks such as PyTorch can conveniently use multiprocessing to request multiple molecules simultaneously
as they are needed in the training process,
freeing the \textit{ParticleGrid} implementation from the need to implement this.
If \textit{ParticleGrid} is indeed memory bound, this will have little impact on performance.
We have yet to fully characterize the impact of multi-threading on the implementation,
but preliminary exploration suggests multi-threading is indeed less effective than would be naively expected. There is more research to be done in this area.
Despite some unknowns in hardware optimization, ParticleGrid is easily mature enough to be part of useful, performant applications.

\section{Application: Property Prediction: Structure is all you need}

\label{sec:application}

It is natural to question whether the information contained in 3D structures is sufficient and required to perform scientific deep learning.  
In order to empirically show the importance of the 3D structure for learning on molecular properties, we train a 3D-CNN to predict molecular properties. 

We perform the training and testing on the hardware mentioned in Table~\ref{tab:cypress config}.
\begin{table}[h]
    \centering
    
    {\def\arraystretch{1.3}
    \begin{tabular}{r|l}
        \toprule
        CPU & AMD EPYC 7F52 (16 cores) / Node \\
        Main Memory & 32 GB / Node \\
        GPU & NVIDIA A100 40 GB (x4) / Node \\
        Compiler & GCC 8.3.1 \\
        ML framework \& Python & Python 3.7 \\ & PyTorch 1.10 \\   & PyTorch-Lightning 1.7.0 \\
        \bottomrule
    \end{tabular}
    \vspace{.2cm}
    \caption{
        Hardware and software configuration for the Cypress cluster, where all of the experiments presented in Section~\ref{sec:application} were performed.
    }
    }
    
    \label{tab:cypress config}
\end{table}

\subsection{Dataset}

We use the  Open Graph Benchmarks Large-Scale Challenge (OGB-LSC) \cite{hu2021ogb} for our exploration on the importance of 3D structure for molecular properties. The PubChem Quantum Mechanics for Molecules (PCQM4M), a part of the OGB-LSC, is a quantum chemistry dataset with about 4 million molecular samples created by the PubChemQC project \cite{pubchemqc, maho2015pubchemqc}. This is a subset of the larger PubChem database, which hosts more than 60 million chemical structures \cite{pubchem}.
The training subset of the OGB-LSC-PCQM4Mv2 dataset, with 3.3 million molecules,  contains 3D coordinate information derived from DFT. Each molecule sample are of varying sizes and a wide array of elements. The dataset sample includes the 3D structures of the molecule and it's accompanying HOMO-LUMO energy gap calculated using DFT. We use 3D coordinates to generate the grids for our model.

The HOMO-LUMO gap is of interest to domain scientists since it describes the chemical and electrical properties of the molecule and is a useful indicator for reactivity and stability. 

Physics-based simulations using DFT are computationally demanding and time-consuming, taking up to several hours for even small molecules.
However, high throughput characterization is important for practical workflows, so efficient machine learning surrogates for DFT simulations have the potential for high impact.

Only the training split of the original dataset contains the 3D structural data, therefore we use that as the basis for our training-validation-test datasets. We perform a 90-5-5 random split of the original dataset. Since our custom dataset splits are not publicly available, we release the code and dataset publicly for reproducibility \footnote{The model and dataset code can be found here: \url{www.github.com/ParticleGrid/ParticleGrid/tree/main/examples/PCQM4M-3D}}.

\begin{figure}[h!]
    \centering
    \includegraphics[width=0.85
    \linewidth]{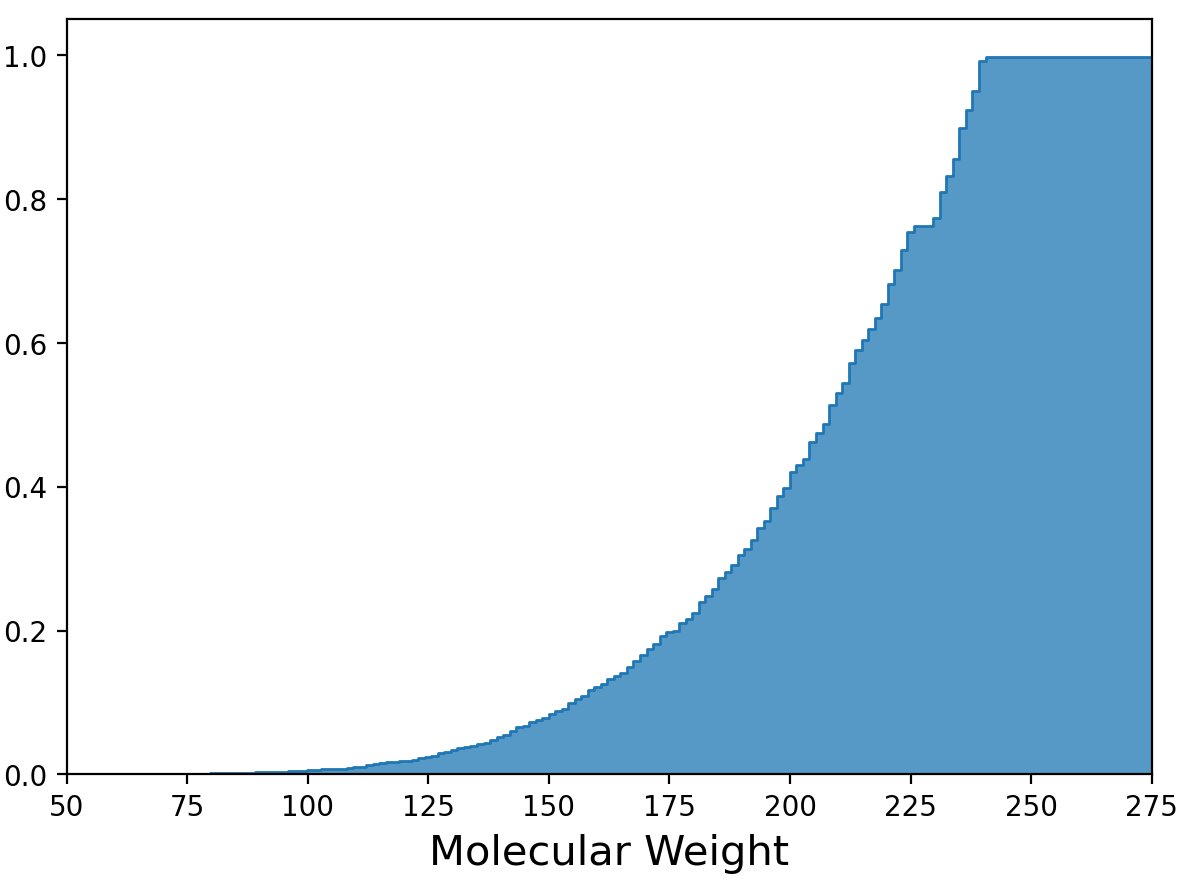}
    \caption{CDF plot of the molecular weight distribution to verify that the dataset has sufficient variety of molecules. }
    \label{fig:weight-dist}
\end{figure}

\subsection{Data Representation}

We are interested in understanding the information stored in the 3D structure of the molecules. Therefore we generate grids where all atoms are treated as the same, and each molecule is represented via a single channel, $(1 \times 32 \times 32 \times 32)$ grid. The grids have adaptive spatial resolution such that the volume covered by each voxel depends on the volume of the bounding box the molecule.  

\subsection{Experiment Details}

A 10-layer 3D-ResNet with skip connections and batch normalization architecture is trained for our experiments. The model is relatively small with only 14.4M trainable parameters. We use the RMSProp optimizer with learning rate $10^{-3}$ and an exponentially decaying learning rate scheduler with $\gamma = 0.999$. During training, we set the batch size to 256 and train concurrently on 4 A100s. We use ParticleGrid with grid size of 32, and variance of 0.4 and generate the grids on the fly during training, validation, and testing. We train for a total of 400 epochs, and save the model with the highest performing validation error. The computation cost is only 166 seconds per epoch, and thus we are able to train the model to completion in less than 20 hours on four A100s. The exact details of the hardware and software stack can be found in Table. \ref{tab:cypress config}.

\subsection{Results}

\begin{figure}[h!]
    \centering
    \includegraphics[width=0.7\linewidth]{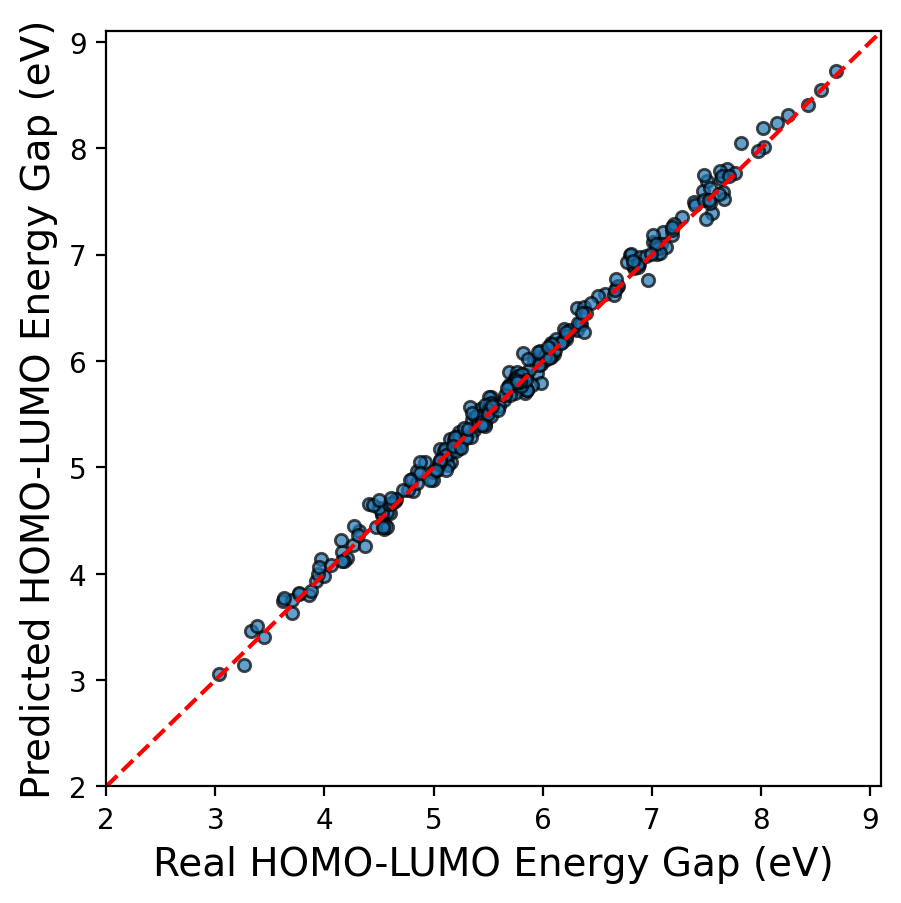}
    \caption{We achieve highly accurate predictions of the HOMO-LUMO energy gap using our 3D-CNN.}
    \label{fig:results}
\end{figure}

We use the mean squared error (MSE) as our metric of evaluation.  We achieve 0.006
MSE on the test dataset. So even though only using the structural information and  single channel grid, we are able to predict the HOMO-LUMO energy gap to DFT-level accuracy. We can see the prediction accuracy in Fig. \ref{fig:results}.  

Furthermore, the computation cost still remains negligible compared DFT. 

As we can see, three dimensional structures contain significant information that can help scientists understand the behavior of the molecules. Significantly, we used no atomic information other than the atomic locations. We can therefore take other molecules and crystals, represent them in the exact representation and use transfer learning for multi-task learning. 

\section{Discussion}

DFT is not the only method to obtain 3D structures materials. Semi-empirical methods such PM and PM7 can provide fast and accurate 3D representation of large scale molecules. Small molecular conformations can be efficiently generated with ETKDG~\cite{ETKDG} with popular cheminformatics tools such as RDKit~\cite{rdkit}.

We do note that our formulation in Section III uses a single element for each element type. While this may be feasible for organic molecules which generally consists of only a few different types elements, it will be inefficient for some materials of interest with a wide variety of possible constituent elements. Since the grid generation process is differentiable, a fact we have used for the discretizer, we can also attach invertible embedding layers to ``learn" molecular embeddings and generate grids using the molecular embeddings. We leave this for future work to enable efficient learnable grid generation on \textit{ParticleGrid}. 

Deep learning kernels for computer vision and natural language processing, such 2D convolutions, dense matrix multiplication have been highly optimized for optimal performance, especially on hardware accelerators, such as GPUs, TPUs, and other ASICs. Optimized GPU kernels in the cuDNN \cite{chetlur2014cudnn} and ROCm libraries enable fast and efficient processing of images and text. 3D layers such convolutions and pooling have not been optimized to the same extent due to utility. Wide scale use of 3D models can spur on hardware and software advancements to further improve performance.  

\section{Conclusion}

We introduce \textit{ParticleGrid} as a middle layer to facilitate deep learning of a wide range of material structures.
The library allows us to generate physics inspired grid representations of molecules and crystals that can be used with deep learning to perform data-driven structure property quantification, or even inverse design. 

Practitioners can take advantage of the gridding functionality to train high-capacity models large scale 3D datasets such as the PubChemQC PM6 \cite{nakata2020pubchemqc} and OGB-LSC-PCQM4M \cite{hu2021ogb}. 
We also propose a transformation to 3D coordinate space to enable applications in generative models and materials discovery. 

ParticleGrid, and the 3D grid-based representations, can be used in conjunction with graph-based and string-based representations to perform predictive and generative modeling of materials.
ParticleGrid provides a simple Python API for ease of use, and to integrate into the data pipelines for deep learning frameworks. 
The CPU optimized algorithm to enable real-time grid generation for deep learning applications. We achieve up to 9000x, 79.5x, and 14x over Numpy, Numba, and baseline C++ implementations, respectively.
Using \textit{ParticleGrid}, we train a 3D CNN to predict molecular HOMO-LUMO energy gap on a dataset of more than 3.3 million in under 20 hours on 4 GPUs, and achieve DFT level accuracy.
We hope fast and efficient 3D grid generating libraries can help further 3D modelling of material structures. 

\section*{Acknowledgements}
We would like to thank TotalEnergies EP Research \& Technology  for supporting and allowing this material to be shared.

\bibliographystyle{IEEEtran}
\bibliography{sample-base}
\end{document}